\documentclass[traditabstract]{aa} 
\usepackage{epsfig}
\usepackage{threeparttable}
\usepackage{graphicx,txfonts,multirow}
\usepackage[T1]{fontenc}
\usepackage{pdflscape}

\def\ltsima{$\; \buildrel < \over \sim \;$}
\def\lsim{\lower.5ex\hbox{\ltsima}}
\def\gtsima{$\; \buildrel $\geq$ \over \sim \;$}
\def\gsim{\lower.5ex\hbox{\gtsima}}
\newcommand{\be}{\begin{equation}}
\newcommand{\en}{\end{equation}}

\def\lum {\mbox{erg s$^{-1}$}}

\newcommand{\axaf}{{\em Chandra}}

\newcommand{\xmm}{{\em XMM--Newton}}
\newcommand{\swift}{{\em Swift}}

\newcommand{\nustar}{{\em NuSTAR}}

\begin{document}

\title{Simultaneous broadband observations and high-resolution X-ray spectroscopy of the transitional millisecond pulsar PSR\,J1023$+$0038}

\author{F. Coti Zelati\inst{1,2,3}, S. Campana\inst{3}, V. Braito\inst{3}, M.~C. Baglio\inst{4,3}, P. D'Avanzo\inst{3}, N. Rea\inst{1,2,5}, and D.~F. Torres\inst{1,2,6}}
\institute{Institute of Space Sciences (ICE, CSIC), Campus UAB, Carrer de Can Magrans s/n, E-08193 Barcelona, Spain
\and
Institut d'Estudis Espacials de Catalunya (IEEC), E-08034 Barcelona, Spain
\and
INAF, Osservatorio Astronomico di Brera, Via E. Bianchi 46, I-23807, Merate (LC), Italy\\
\email{cotizelati@ice.csic.es}
\and
New York University Abu Dhabi, P.O. Box 129188, Abu Dhabi, United Arab Emirates
\and
Anton Pannekoek Institute for Astronomy, University of Amsterdam, Postbus 94249, NL-1090-GE Amsterdam, The Netherlands
\and
Instituci\'o Catalana de Recerca i Estudis Avan\c{c}ats (ICREA), E-08010 Barcelona, Spain}

\date{Received 6 November 2017 / Accepted 30 December 2017}

\abstract{We report on the first simultaneous \xmm, \nustar,\ and \swift\ observations of the transitional millisecond pulsar PSR\,J1023$+$0038 
in the X-ray active state. Our multi-wavelength campaign allowed us to investigate with unprecedented detail possible spectral variability over 
a broad energy range in the X-rays, as well as correlations and lags among emissions in different bands. The soft and hard X-ray emissions are 
significantly correlated, with no lags between the two bands. On the other hand, the X-ray emission does not correlate with the UV emission. 
We refine our model for the observed mode switching in terms of rapid transitions between a weak propeller regime and a rotation-powered radio 
pulsar state, and report on a detailed high-resolution X-ray spectroscopy using all \xmm\ Reflection Grating Spectrometer data acquired since 2013. 
We discuss our results in the context of the recent discoveries on the system and of the state of the art simulations on transitional millisecond pulsars, 
and show how the properties of the narrow emission lines in the soft X-ray spectrum are consistent with an origin within the accretion disc.}

\keywords{methods: data analysis --  methods: observational -- stars: magnetic field -- stars: neutron -- pulsars: individual: PSR J1023+0038 -- X-rays: binaries.}

\titlerunning{Simultaneous multiband observations and high-resolution X-ray spectroscopy of J1023}
\authorrunning{F. Coti Zelati et al.} 

\maketitle

\section{Introduction}

According to the pulsar recycling scenario first proposed in the 1980s (Alpar et al. 1982; Radhakrishnan \& Srinivasan 1982; Taam 
\& van den Heuvel 1986; Bhattacharya \& van den Heuvel 1991; see also Tauris \& van den Heuvel 2006; Srinivasan 2010), 
millisecond radio pulsars (MSPs) are old pulsars that have been spun up to short spin periods after a Gyr-long phase 
of mass transfer in a low-mass X-ray binary system (LMXB). 
Repeated swings between an accretion-disk-dominated LMXB state and a rotation-powered MSP state were recently observed 
over a few weeks in the system IGR\,J18245$-$2452 (Papitto et al. 2013). This discovery demonstrated that these transitions are 
not an irreversible, long-lasting process, but can also occur on time scales compatible with those of the variations of the mass 
accretion rate onto the neutron star (NS). At the moment of writing, the class of these so-called transitional millisecond pulsars (TMPs) 
includes three confirmed members: IGR\,J18245$-$2452 (Papitto et al. 2013), PSR\,J1023$+$0038 (Archibald et al. 2009) and 
XSS\,J12270$-$4859 (Bassa et al. 2014).

Observed as an eclipsing 1.69-ms radio pulsar in a 4.75-h binary orbit between 2007 and 2013 (Archibald et al. 2013), 
PSR\,J1023$+$0038 (J1023) has been the subject of extensive multi-wavelength monitoring campaigns since 2013 June, 
when radio pulsations became undetectable at all orbital phases (Stappers et al. 2014), and double-peaked H$\alpha$ 
emission lines -- the signature of an accretion disc -- were clearly observed in the optical spectrum (Halpern et al. 2013). 

Radio-timing monitoring programs with different facilities failed to detect pulsations at the NS spin period following 
the transition to the accretion state (see Jaodand et al. 2016 and references therein). Imaging observations in the 
1--18~GHz frequency range between 2013 November and 2014 April unveiled a highly variable emission on timescales 
as short as minutes, with a flat spectrum possibly related to synchrotron emission originating in material outflowing from 
the system (e.g. in the form of a weakly collimated partially self-absorbed jet; Deller et al. 2015). The gamma-ray 
luminosity of J1023 quintupled approximately at the same epoch at which the radio pulsations disappeared, and enhanced 
0.1--300~GeV emission has been recorded since then (Stappers et al. 2014; Deller et al. 2015; see also Torres et al. 2017 
for a more detailed study). A search for steady emission above 300~GeV in data acquired in 2013 December with the 
\emph{VERITAS} gamma-ray observatory resulted instead in an upper limit for the luminosity $L_{>300~{\rm GeV}} 
< 1.5 \times 10^{32}$~\lum\ (adopting the parallax distance of 1.37~kpc calculated by Deller et al. 2012) at the 95 
percent confidence level (Aliu et al. 2016).

A \nustar\ observation in 2013 October led to an estimated luminosity of about $6 \times 10^{33}$~\lum\ in the 3--79~keV band, 
a factor of $\sim8$ larger than that attained in the rotation-powered MSP state (Li et al. 2014; Tendulkar et al. 2014). Evidence 
for significant X-ray variability was detected in the \swift\ data sets on time scales of a few tens of seconds (Patruno et al. 2014), 
but also of days/weeks (Coti Zelati et al. 2014; Takata et al. 2014). However, a set of nine observations carried out with the larger 
collecting area of the X-ray instruments on board \xmm\ between 2013 November and 2015 December have all revealed a puzzling 
trimodal behaviour of the system in the soft X-rays (0.3--10~keV): J1023 spends about 70--80 percent of the time in a stable `high' 
mode ($L_X \sim 3.4 \times 10^{33}$~\lum), which unpredictably alternates to a `low' mode ($L_X \sim 5 \times 10^{32}$~\lum). 
Sporadic flaring episodes are also observed -- although not in all the observations -- reaching luminosities $L_X \sim 10^{34}$~\lum\ 
(e.g. Jaodand et al. 2016). The switches between these modes occur on a timescale of tens of seconds. Coherent X-ray pulsations 
at the NS spin period are detected only when the system is in the high mode (Archibald et al. 2015), and the phase-connected timing 
solution reported by Jaodand et al. (2016) shows that the average NS spin-down rate in the LMXB state is only about 27 percent 
larger than that observed during the rotation-powered MSP state. This strongly suggests not only that the pulsar spin-down mechanism 
has been operating since the formation of an accretion disc around the NS, but also that the outflow of material has increased the 
spin-down and largely prevails over the spin-up torque imparted by the accreting material.
   
Recent simultaneous \axaf\ and \emph{Karl G. Jansky Very Large Array} observations showed an anti-correlated variability 
pattern between the X-ray and radio emissions: the X-ray low mode is accompanied by a temporally coincident radio brightening, 
while the X-ray high mode occurs in correspondence of steady low-level radio emission (Bogdanov et al. 2017).

The system became $\sim3.5$ and $\sim1$ mag brighter in the UV and optical bands, respectively, compared to the rotation-powered 
MSP state (Halpern et al. 2013; Patruno et al. 2014). Optical photometry revealed modulation of the emission at the 4.75-h orbital 
period, (Coti Zelati et al. 2014; McConnell et al. 2015; Shahbaz et al. 2015; Papitto et al., in preparation), due to reprocessing of the X-ray 
emission by the heated face of the companion star (Papitto et al., in preparation). Flaring activity with rapid ($\lesssim 3$~s) transition 
timescales was observed both in the optical and near infrared bands (Hakala \& Kajava 2018). The first optical polarimetric study of J1023 
was conducted by Baglio et al. (2016), who suggested that the measured linear polarization of $\sim1$ percent in the system emission and 
its possible modulation at the orbital period could be due to Thomson scattering with electrons in the disc. This would exclude the presence 
of a well-structured jet at optical wavelengths. More recently, Hakala \& Kajava (2018) unveiled a significant anticorrelation between the 
optical flux and the degree of linear polarization. They also reported on the possible presence of an additional transient polarized emission 
component during the flares, and deduced changes in the disc emission during these episodes from H$\alpha$ spectroscopy and Doppler 
tomography. These results led the authors to propose that Thomson scattering from matter ejected by the propelling magnetosphere might 
be the mechanism powering the optical flares. Optical pulsations at the NS spin period were detected in observations carried out in 2016 March, 
making J1023 the first optical MSP ever detected (Ambrosino et al. 2017). This discovery led the authors to favour a scenario in which the pulsed 
emission originates from synchrotron emission by relativistic electrons and positrons in the magnetosphere of a rotation-powered pulsar.

The wealth of data collected for J1023 has hence revealed unique properties of the system, and finding a unified scenario able of 
taking into account all of the above phenomenological characteristics of this system, as well as of the other transitional pulsars, has represented 
a tough challenge in the last years. In a previous study we interpreted the repeated transitions between the high and low modes in terms 
of fast swings between the propeller regime and the rotation-powered MSP state (Campana et al. 2016; see also Linares 2014), and 
showed how this scenario could satisfactorily explain the multiwavelength properties of J1023 observed over the past 4 years. Here we present 
new simultaneous \xmm, \nustar,\ and \swift\ observations which allowed us to search for correlations and lags between multiband emissions,
as well as to improve our model for the observed mode switching of the system. We also present the results of our searches for narrow spectral 
features in archival Reflection Grating Spectrometer data with the aim of constraining the location of the emitting regions. 

The study is structured as follows: we describe our multi-wavelength observations and the data reduction in Sect.~\ref{reduction}. We 
show the results of our searches for correlations and lags among the different energy bands in Sect.~\ref{crosscor}. We perform the first strictly 
simultaneous broadband characterisation of the different modes and refine our modelling for the mode switching in Sect.~\ref{model}. 
We present a detailed characterisation of the high-resolution X-ray spectrum in Sect.~\ref{sec:rgs_results}. Discussion of the results and 
conclusions follow in Sect.~\ref{discussion}.

\begin{table*}
\begin{center}
\caption{Log of the simultaneous observations of J1023 on 2016 May 8--9.} 
\label{tab:log}
\resizebox{2.\columnwidth}{!}{
\begin{tabular}{ccccccc} \hline \hline
Satellite                               & Obs. ID                                       & Instrument                                      & Mode                                  & Start -- End time of exposure                                           & Exposure$^{\rm a}$ & Source net count rate$^{\rm b}$ \\
                                        &                                               &                                                         &                                                & (UTC)                                                                         & (ks)                    & (counts~s$^{-1}$)\\ 
\hline 
\vspace{0.15cm}
\multirow{8}{*}{\xmm}   & \multirow{8}{*}{0784700201}   & EPIC pn                                               & FT                                              & 2016/05/08 04:21:57 -- 2016/05/09 14:46:27                     & 121.7                         & $3.641 \pm 0.008$\\ 
                                        &                                               & EPIC MOS\,1                                     & SW                                    & 2016/05/08 03:47:12 -- 2016/05/09 12:08:48                      & 112.9                         & $1.001 \pm 0.003$\\ \vspace{0.15cm}
                                        &                                               & EPIC MOS\,2                                     & SW                                    & 2016/05/08 03:47:57 -- 2016/05/09 12:09:02                      & 113.0                         & $1.004 \pm 0.004$\\                             
                                        &                                               & RGS\,1                                          & \multirow{2}{*}{Spectroscopy} & 2016/05/08 03:47:02 -- 2016/05/09 14:42:33                      & 125.4                          & \multirow{2}{*}{$0.0893 \pm 0.0008$} \\ \vspace{0.15cm}                       
                                        &                                               & RGS\,2                                          &                                               & 2016/05/08 03:47:10 -- 2016/05/09 14:43:21              & 125.6                         & \\      
                                        &                                               & OM                                              & Image                                  & 2016/05/08 07:49:52 -- 2016/05/09 16:46:19                    & 100.0                   & -- \\
\hline
\multirow{2}{*}{\nustar}        & \multirow{2}{*}{30201005002}  & FPMA                                          & \multirow{2}{*}{Science}                & \multirow{2}{*}{2016-05-07 15:34:34 -- 2016-05-09 15:13:26}         & 86.0          & $0.173 \pm 0.001$ (70\%)\\
                                        &                                               & FPMB                                            &                                               &                                                                                       & 85.7                    & $0.166 \pm 0.001$ (71\%)\\
\hline 
\swift                                  & 0790180101                            & UVOT                                            & event                                 & 2016/05/08 18:45:01 -- 2016/05/09 12:21:57      & 9.5                           & $0.90 \pm 0.01$\\                       
\hline                                          
\end{tabular}
}
\end{center}
{\bf Notes.}
\noindent $^{\rm a}$ Deadtime corrected on-source exposure.\\ 
\noindent $^{\rm b}$ Background-subtracted count rate in the 0.3--10~keV energy band for the EPIC instruments, in the 0.35--2.5~keV energy band for the combined RGS first order data sets,  
in the 3--79~keV energy band for the \nustar\ data (the number in parentheses representing the fractional contribution from the 3--10~keV band), and in the $UVM2$ filter for the \swift\ UVOT data.
\end{table*}

\section{Observations and data extraction}
\label{reduction}

\subsection{XMM-Newton}
\label{xmm}

J1023 was observed with the European Photon Imaging Cameras (EPIC) on board \xmm\ (Jansen et al. 2001) 
starting on 2016 May 8 at 03:29 UT and over one entire revolution (see Table~\ref{tab:log} for details on 
the observation). The pn (Str\"uder  et al. 2001) was set in fast timing (FT) mode, where data are read 
every 29.52~$\mu$s and only one-dimensional imaging is preserved. The two MOS cameras (Turner et al. 
2001) were operated in small window mode (SW, 0.3-s time resolution) over the first 113~ks of the 
observation. During the last 6.7-ks segment of the observation, they both switched to a different observational 
setup and experienced serious telemetry issues, yielding a dramatic and unphysical drop of the source flux. 
In the following, we do not consider these data sets, and focus only on the longest uninterrupted exposures 
covering the first 113~ks of the observation. The thin optical blocking filter was positioned in front of the three 
cameras. The Reflection Grating Spectrometer (RGS; den Herder et al. 2001) arrays  were configured in standard 
spectroscopy mode (5.7-s time resolution) throughout the observation. The Optical/UV Monitor Telescope (OM; 
Mason et al. 2001) collected the data both in the `Image' and `Fast Window' modes, and used the $B$ filter, 
which is centred at 4392~\AA\ and covers the wavelength interval from 3800 to 5000~\AA. Unfortunately, J1023 
fell outside the small window of the instrument in the `Fast Window' mode data sets due to a small attitude problem 
(N. Schartel, priv. comm.). We thus focus on the image mode data alone. A total of 27 exposures were 
acquired, resulting in a total deadtime-corrected on source exposure time of 100~ks.

\begin{figure*}
\centering
\includegraphics[width=1\textwidth]{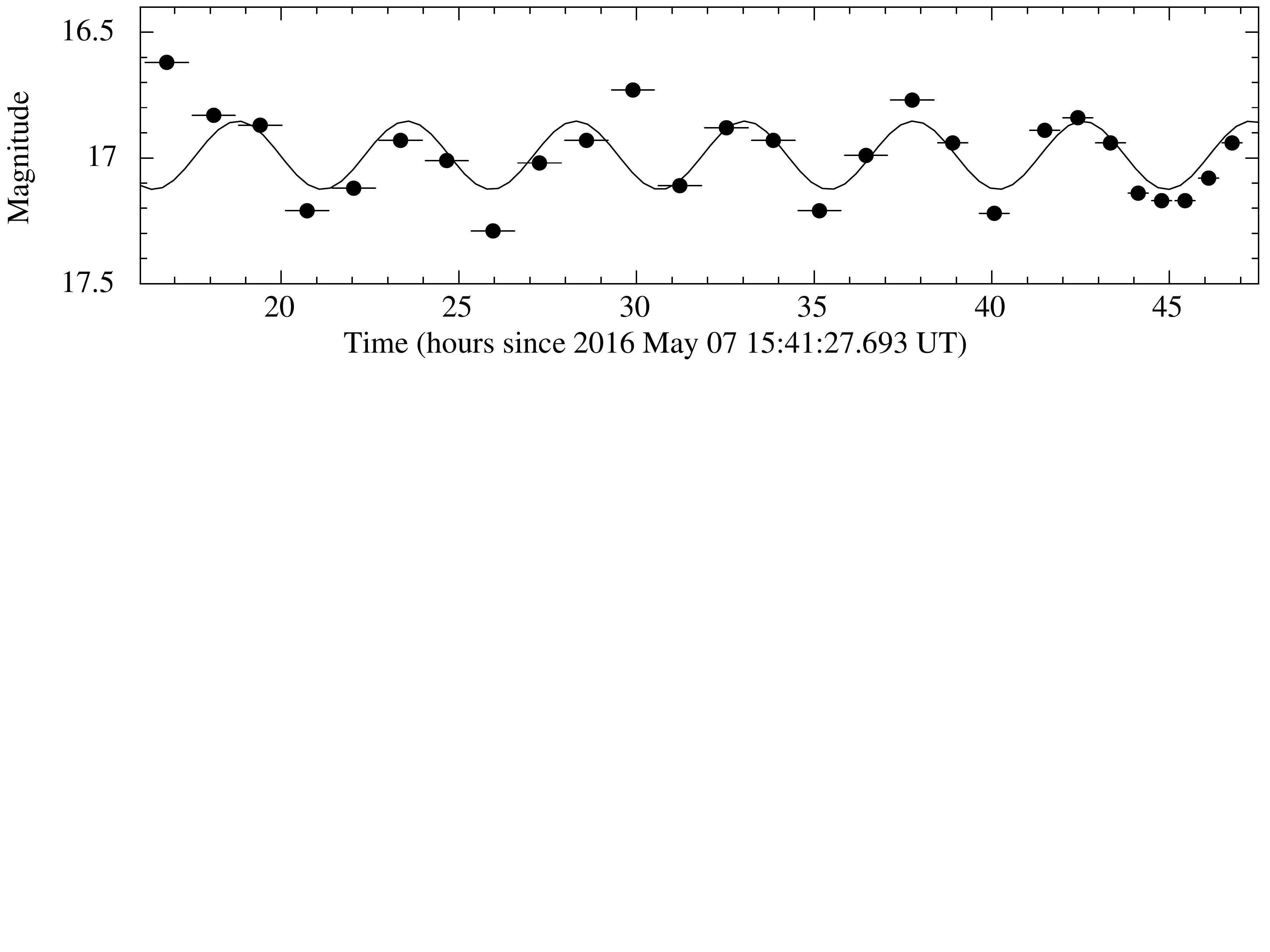}
\vspace{-8.8cm}
\caption{Optical light curve of J1023 in the $B$ band. The solid line represents the best-fitting sinusoidal function 
(the period was fixed at the orbital period reported by Jaodand et al. (2016); see the text for details). Errors on the 
magnitudes are smaller than the size of the markers.}
\label{fig:om_lcurve}
\end{figure*}

\subsubsection{Optical/UV Monitor Telescope data}
\label{om}

We reduced the data using the \textsc{omichain} processing pipeline with the default parameter settings, as 
recommended by the \textsc{sas} threads. We determined the source magnitudes (reddened and in the Vega 
system) with \textsc{omsource} to extract the optical light curve, which is shown in Fig.~\ref{fig:om_lcurve}. 
A fit with a sinusoidal function with period fixed to the orbital period of the binary system (17716~s; see Jaodand 
et al. 2016) yields $\chi^2_\nu = 205.65$ for 23 degrees of freedom (dof), a result that clearly hints at additional 
variability superimposed on any orbital modulation, possibly in the form of flaring episodes as in the X-rays. 
Unfortunately, a detailed study of this variability on shorter timescales (similarly to what is reported by Jaodand et al. 
2016) is precluded since the fast-window mode data do not provide any useful information on the source because 
of the above mentioned loss of data.

\subsubsection{EPIC data}

We processed the raw observation data files using the \xmm\ Science Analysis System 
(\textsc{sas}\footnote{See http://www.cosmos.esa.int/web/xmm-newton/sas.} 
v. 15.0) and the most up to date calibration files available (XMM-CCF-REL-332).

We built a light curve of single-pixel events (\textsc{pattern} = 0) for the whole field of view and 
for each camera, to visually inspect for the presence of enhanced background flaring activity 
induced by soft-proton flares of solar origin. We revealed no significant contamination in all the 
three cameras throughout the observation, except for the last 10~ks in the pn data sets, which 
were hence excluded from the following analysis.

For the pn camera we extracted the source counts within a 10-pixel-wide strip centred on the 
brightest column of CCD pixels (RAW-X = 37), and oriented along the readout direction of the CCD. 
Background events were accumulated within a similar region sufficiently far from the position of the 
source so as to minimize the contribution from the wings of the point spread function (PSF). For the 
two MOSs, source counts were collected within a circle centred on the position of the radio counterpart 
of the system ($\rm RA=10^h23^m47^s.69$, $\rm Dec= 00^\circ38'40''.85$; Archibald et al. 2013) and 
with a radius of 36 arcsec. The background was extracted from a circle of the same size located on the 
same CCD and far from detector areas possibly contaminated by out-of-time events from the source or 
too near to the CCD edges. Photon arrival times of all event lists were referred to the solar system 
barycentre reference frame using \textsc{barycen} and the DE-200 solar system ephemeris (which we 
adopt in the following for all data sets).

\begin{table*}
\begin{center}
\caption{Log of the \xmm\ RGS observations of J1023 since the system transition to the X-ray active state.} 
\label{tab:rgs_log}
\begin{tabular}{ccccc} \hline \hline
Obs. ID                                         & Instrument                    & Start -- End time of exposure                                   & Exposure         & Source net count rate \\
                                                &                               & (UTC)                                                                   & (ks)            & (counts~s$^{-1}$)\\ 
\hline
\multirow{2}{*}{0720030101}     & RGS\,1                        & 2013/11/10 17:21:55 -- 2013/11/12 06:03:36                 & 131.8                  & $0.079\pm0.001$ \\ \vspace{0.15cm}                    
                                                & RGS\,2                        & 2013/11/10 17:29:01 -- 2013/11/12 06:03:37              & 131.5          & $0.092\pm0.001$ \\
\multirow{2}{*}{0742610101}     & RGS\,1                        & 2014/06/10 03:44:53 -- 2014/06/11 12:41:17         & 116.5         & $0.074\pm0.001$ \\ \vspace{0.15cm}                      
                                                & RGS\,2                        & 2014/06/10 03:45:01 -- 2014/06/11 12:42:07              & 116.6         & $0.084\pm0.001$ \\
\multirow{2}{*}{0748390101}     & RGS\,1                        & 2014/11/21 18:32:36 -- 2014/11/22 04:09:25                 & 34.5          & $0.091\pm0.002$ \\ \vspace{0.15cm}                      
                                                & RGS\,2                        & 2014/11/21 18:32:44 -- 2014/11/22 04:09:30              & 34.5          & $0.105\pm0.002$ \\
\multirow{2}{*}{0748390501}     & RGS\,1                        & 2014/11/23 18:24:42 -- 2014/11/24 04:09:55         & 35.0                  & $0.073\pm0.002$ \\ \vspace{0.15cm}                      
                                                & RGS\,2                        & 2014/11/23 18:24:50 -- 2014/11/24 04:09:57              & 35.0          & $0.082\pm0.002$ \\
\multirow{2}{*}{0748390601}     & RGS\,1                        & 2014/11/28 21:31:11 -- 2014/11/29 03:19:47         & 20.8                  & $0.084\pm0.003$ \\ \vspace{0.15cm}                      
                                                & RGS\,2                        & 2014/11/28 21:31:18 -- 2014/11/29 03:19:42              & 20.8          & $0.095\pm0.003$ \\
\multirow{2}{*}{0748390701}     & RGS\,1                        & 2014/12/17 15:45:29 -- 2014/12/18 01:24:06         & 34.6                  & $0.080\pm0.002$ \\ \vspace{0.15cm}                      
                                                & RGS\,2                        & 2014/12/17 15:45:37 -- 2014/12/18 01:24:01              & 34.5          & $0.091\pm0.002$ \\
\multirow{2}{*}{0770581001}     & RGS\,1                        & 2015/11/11 19:36:35 -- 2015/11/12 03:48:28         & 29.4                  & $0.081\pm0.002$ \\ \vspace{0.15cm}                      
                                                & RGS\,2                        & 2015/11/11 19:36:43 -- 2015/11/12 03:48:29              & 29.4          & $0.097\pm0.002$ \\
\multirow{2}{*}{0770581101}     & RGS\,1                        & 2015/11/13 03:30:32 -- 2015/11/13 09:52:31         & 22.8                  & $0.120\pm0.003$ \\ \vspace{0.15cm}                      
                                                & RGS\,2                        & 2015/11/13 03:30:40 -- 2015/11/13 09:52:31              & 22.8          & $0.134\pm0.003$ \\
\multirow{2}{*}{0783330301}     & RGS\,1                        & 2015/12/09 01:03:46 -- 2015/12/09 08:27:21         & 26.5                  & $0.084\pm0.002$ \\ \vspace{0.15cm}                      
                                                & RGS\,2                        & 2015/12/09 01:03:53 -- 2015/12/09 08:27:20              & 26.5          & $0.097\pm0.002$ \\
\multirow{2}{*}{0784700201}     & RGS\,1                        & 2016/05/08 03:47:02 -- 2016/05/09 14:42:33         & 125.4                 & $0.083\pm0.001$ \\ \vspace{0.15cm}                      
                                                & RGS\,2                        & 2016/05/08 03:47:10 -- 2016/05/09 14:43:21      & 125.6         & $0.096\pm0.001$ \\
\hline                                          
\end{tabular}
\end{center}
{\bf Notes.} The RGS instruments were configured in the Spectroscopy mode in all observations. Count rates are quoted for the 0.35--2.5~keV energy band.
\end{table*}

\subsubsection{RGS data}
\label{rgs}

We processed the data of each RGS instrument using the task \textsc{rgsproc}, and produced calibrated 
and concatenated photon event lists, source and background spectra, and response matrices. We reduced 
the data also for all the other nine observations performed during the LMXB state (see Table~\ref{tab:rgs_log} 
for a journal of these observations), and focused our analysis on the first-order spectra alone, which have a 
higher number of counts. A total of about 134\,500 counts were collected by both RGSs in an exposure time 
of $\sim580$~ks. 

The X-ray spectral properties of J1023 are remarkably similar among all \xmm\ observations, both averaged 
(see e.g. Bogdanov et al. 2015) and in the different X-ray modes (Campana et al. 2016). We assumed that the 
line properties do not vary among the observations, and combined all background-subtracted spectra of each 
RGS module using the \textsc{rgscombine} script. We then verified that the stacked spectra of each RGS module 
were in good agreement with each other, and combined them into a single spectrum (\textsc{rgscombine} 
appropriately accounts for the response matrices and backgrounds of the different spectrometers). We then rebinned 
the spectrum by a factor of 5 (corresponding to the full-width at half-maximum of the instrument), and limited the 
analysis to the 6.2--35.4~\AA\ wavelength interval (0.35--2~keV), where the first-order spectra are best calibrated. 
We verified that rebinning the spectrum by a factor of 3 did not yield different values for the line parameters.

\subsection{NuSTAR}
\label{nustar}

The \emph{Nuclear Spectroscopic Telescope Array} mission (\nustar; Harrison et al. 2013) pointed its 
mirrors towards J1023 starting on 2016 May 7 at UT 15:34:34 and for a total net exposure time of about 
86~ks (see Table~\ref{tab:log}). We processed the event files using the script \textsc{nupipeline} (v. 0.4.5) 
of the \nustar\ Data Analysis Software (\textsc{nustardas}, v.1.6.0, distributed along with \textsc{heasoft} 
v. 6.20), and the instrumental calibration files stored in \textsc{caldb} v20160824. 
As recommended in the online \nustar\ Data Analysis Software Guide\footnote{See https://heasarc.gsfc.nasa.gov/docs/nustar/analysis/nustar$_-$sw\-guide.pdf.}, 
we cleaned the unfiltered event files with the standard depth correction, significantly reducing the internal 
high-energy background. We also removed passages through the South Atlantic Anomaly setting 
`\textsc{saamode}=optimized'. We adopted the latest version of the \nustar\ clock file available (v. 66) to 
correct for drifts of the spacecraft clock caused by changing thermal conditions within the spacecraft. 
We collected the source counts within a circular region of radius 50 arcsec, and background photons from 
a circle of radius 135 arcsec located as far away from the source as possible and on the same detector chip. 
This approach induces small systematic uncertainties in the background, which is known to change from 
chip to chip (e.g. Wik et al. 2014). Corrections to the photons' times of arrival were applied by means of the
 \textsc{barycorr} tool.

\subsection{Swift UVOT}
\label{swift}

The \swift\ satellite observed J1023 for about 9.5 ks, starting on 2016 May 8 at 18:44:56 UT. In this study we focus 
on data acquired with the Ultraviolet and Optical Telescope (UVOT; Roming et al. 2005), since the energy band 
covered by the X-ray Telescope (XRT; Burrows et al. 2005) is fully covered by the EPIC cameras and the XRT provide 
a much poorer counting statistics, a coarser time resolution (2.5~s) and uneven sampling of the source variability 
with respect to the \xmm\ data. 

The UVOT was operated in event mode, yielding a minimum time resolution of about 11~ms, and used the $UVM2$ 
filter (central wavelength of 2246~\AA; see Poole et al. 2008 and Breeveld et al. 2011 for more details on the filters). 
We converted the raw coordinate positions into detector and sky coordinates and removed hot pixels with the 
\textsc{coordinator} and \textsc{uvotscreen} scripts, respectively. We then extracted the source and background counts 
within \textsc{xselect}, using a circle of radius 5 arcsec to collect the source photons and a close-by circle of radius 20 
arcsec for the background. Photon arrival times were referred to the solar system barycentre reference frame using 
the \textsc{barycorr} task and the satellite orbit ephemeris file.

\begin{figure*}
\hspace{+10cm}
\includegraphics[width=1.28\textwidth]{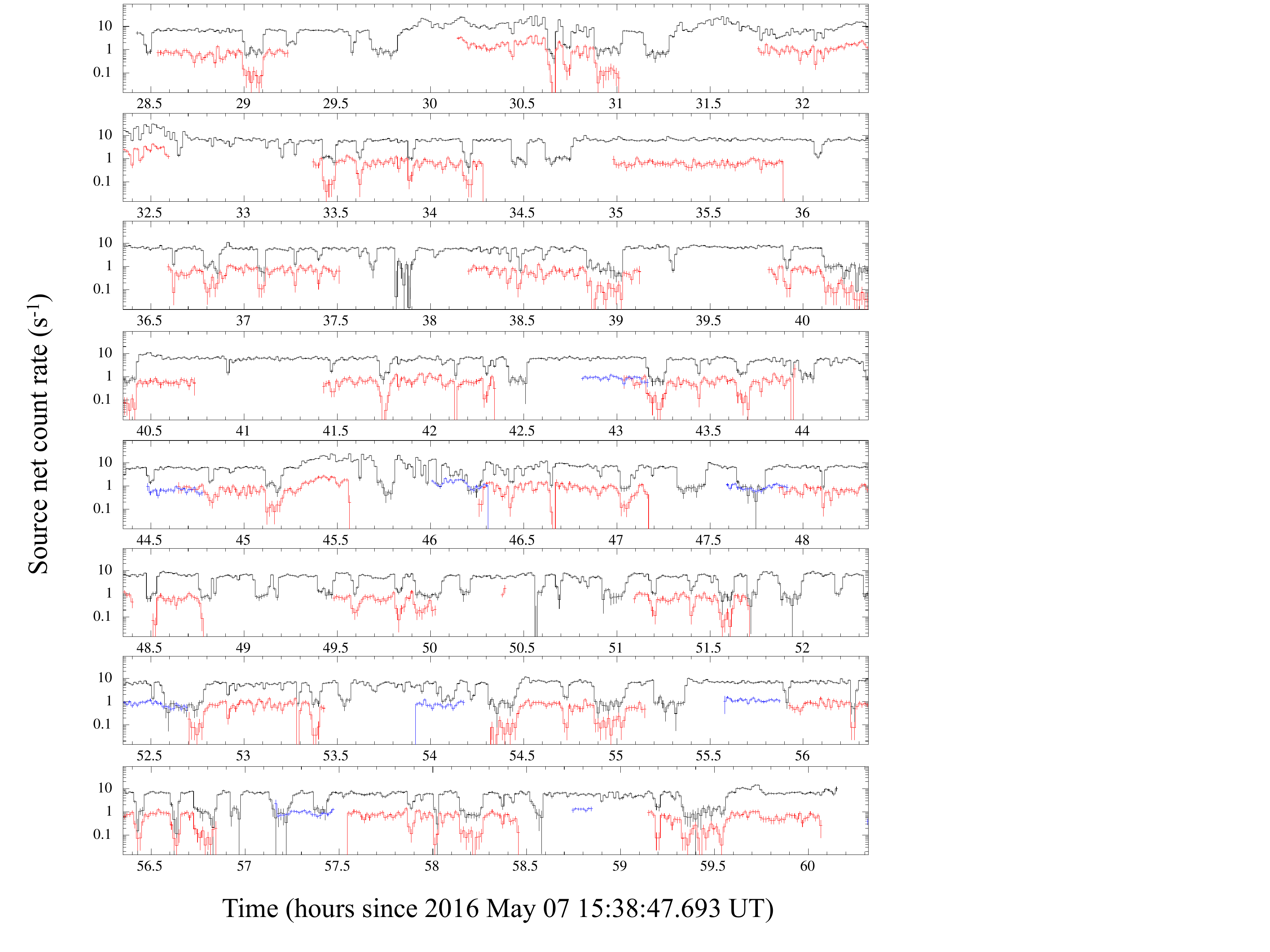}
\centering
\includegraphics[width=.48\textwidth]{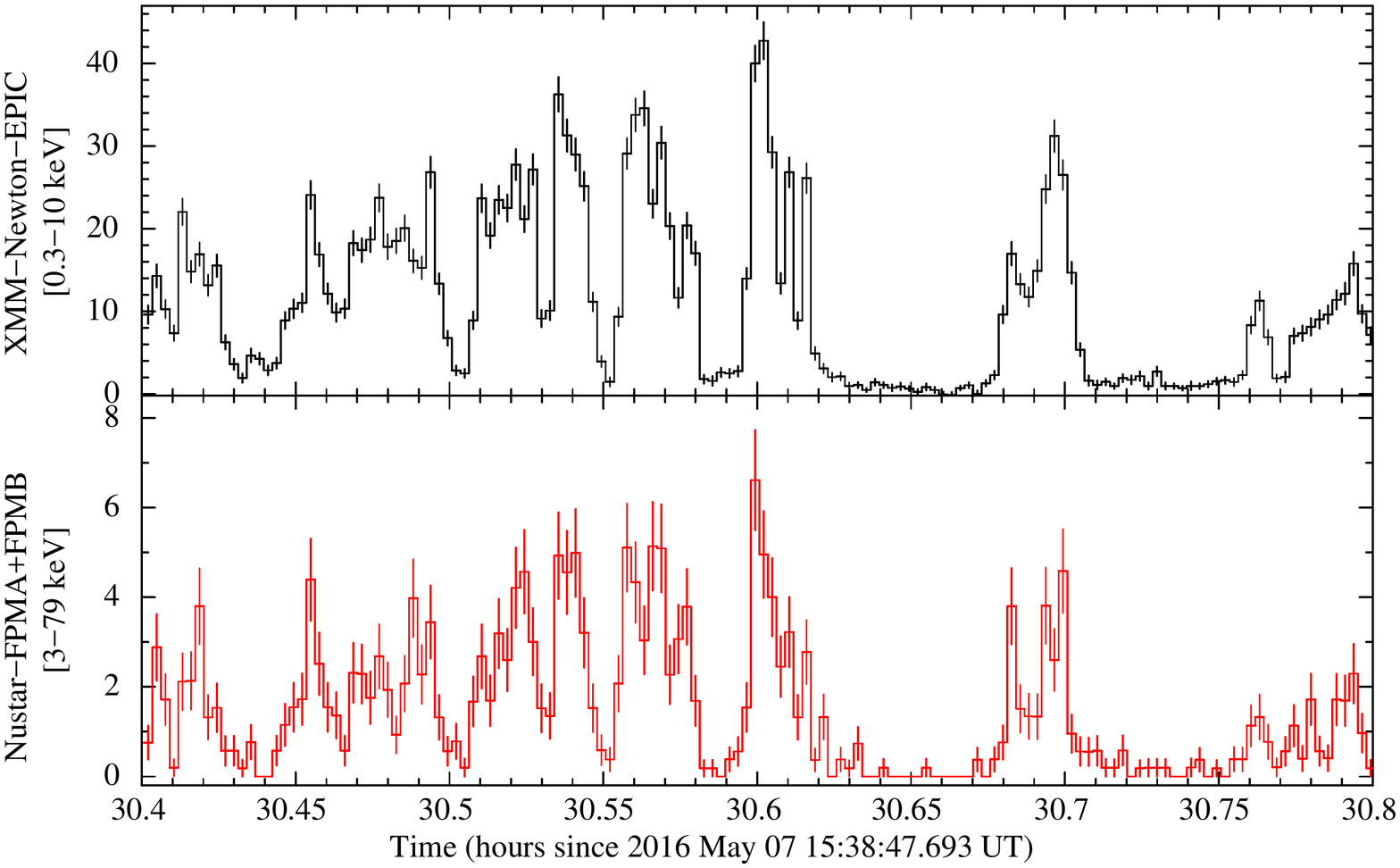}
\vspace{-1.2cm}
\caption{Top: background-subtracted and exposure-corrected light curves of J1023 obtained with the \xmm\ EPIC cameras 
(0.3--10~keV; black data), \nustar\ FPMA + FPMB (3--79~keV; red data) and \swift\ UVOT (UVM2 filter; blue data) during 
the time interval covered by \xmm. For plotting purposes, light curves are shown with a binning time of 50~s and the vertical 
axis is plotted in logarithmic scale. Bottom: zoom of the  \xmm\ EPIC and \nustar\ FPMA + FPMB light curves around the 
longest and brightest flaring episode detected in the data. Light curves are shown with a binning time of 10~s and the vertical 
axis is plotted in linear scale.}
\label{fig:xmm_lcurve}
\end{figure*}

\section{Light curves, correlations, and lags}
\label{crosscor}

The light curves of J1023 over the time interval covered by the \xmm\ observation are reported in 
Fig.~\ref{fig:xmm_lcurve}. Black, red, and blue data points refer to the \xmm\ EPIC, \nustar\ FPMA+FPMB 
and \swift\ UVOT data sets, respectively. We first extracted the 0.3--10~keV light curve from the \xmm\ EPIC 
data sets by combining the 10-s binned background-subtracted and exposure-corrected time series from each 
EPIC camera during the periods when all three telescopes acquired data simultaneously. Corrections for bad 
pixels, chip gaps, PSF variation, vignetting, quantum efficiency and dead time were also applied by means of 
the \textsc{epiclccorr} tool. We then built the 3--79~keV light curves from \nustar\ data separately for FPMA 
and FPMB using the \textsc{nuproducts} script, and combined them to increase the signal-to-noise ratio. Light 
curves were also extracted separately in the 3--10~keV and 10--79~keV energy bands. We then extracted the 
UV exposure-corrected and background-subtracted light curve for the \swift\ event lists by applying the 
\textsc{lcmath} task, accounting for different areas of the source and background extraction regions.

The \xmm\ and \nustar\ light curves display rapid and unpredictable variations in the count rate that are similar  
in all respects to those observed in all the previous \xmm\ observations of the system while in its current state 
(Fig.~\ref{fig:xmm_lcurve}; see also Bogdanov et al. 2015; Jaodand et al. 2016). For the \xmm\ time series 
we found that, over the 32~h of the observation, the system spends about 3 percent of its time in a flaring mode 
(which we define for consistency with previous works as the mode where the source net count rate in the EPIC light
curve binned at 10~s exceeds 15 counts~s$^{-1}$), $\sim 66$ percent of the time in the `pure' high mode (count 
rate between 4 and 11 counts~s$^{-1}$), and $\sim 17$ percent of the time in the `pure' low mode (when the count 
rate drops below 2.1 counts~s$^{-1}$). In particular, we observe a sequence of three flaring events in the time series 
(between $\sim29.8$~h and $\sim32$~h in Fig.~\ref{fig:xmm_lcurve}), followed by two additional episodes 
(between $\sim45.3$~h and $\sim46.1$~h in Fig.~\ref{fig:xmm_lcurve}). 
During the brightest flares, the count rate in the EPIC light curves reaches values as high as $\sim40$ counts~s$^{-1}$. 
The fractions of the time that the system spends in the different modes are remarkably similar to those observed in the
other two deep \xmm\ observations (Archibald et al. 2015; Bogdanov et al. 2015). We also note that J1023 spends a 
non-negligible ($\sim 14$ percent) amount of time in an intermediate state between the `pure' high and low modes, 
related to the periods where the system switches between the two modes and during which the count rate attains a 
value outside the boundaries defined above. We computed the hardness ratio between soft (0.3--10~keV) and hard 
(10--79~keV) X-ray bands, and found no substantial variation of the spectral shape across all modes.
 
\begin{figure}
\hspace{-0.5cm}
\includegraphics[width=.56\textwidth]{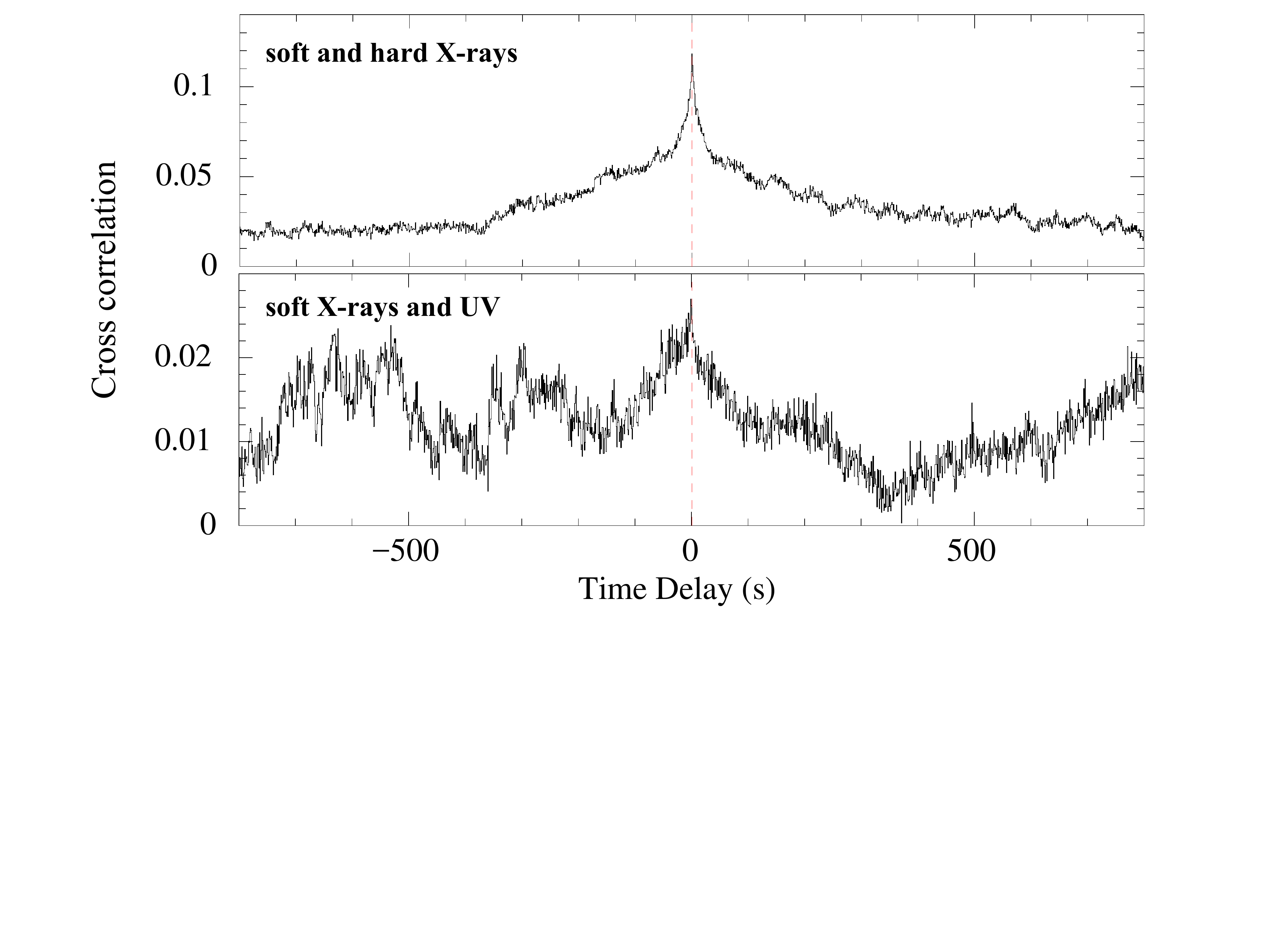}
\vspace{-3cm}
\caption{Results of the cross-correlation between the time series in the 0.3--10~keV and 10--79~keV energy bands 
(\emph{top panel}), and in the UV and the 0.3--10~keV bands (\emph{bottom panel}). In both panels lags are defined 
in such a way that positive time delays refer to the pn data with respect to FPM data (top), and to the pn data with respect 
to the UVOT data (bottom). In both panels the red dashed line marks the case of no time lag. We note the different interval on the vertical axis.}
\label{fig:crosscor}
\end{figure}

We investigated possible time lags (on different timescales, from 0.1~s up to a few seconds) among the emission 
in the different energy bands using the task \textsc{crosscor} of \textsc{Xronos}. The shape of the cross-correlation 
function between the soft X-ray (0.3--10~keV, from \xmm\ pn data) and hard X-ray (10--79~keV, from \nustar\ 
FPMA+FPMB data) is clearly symmetrical and shows no lags on the above-mentioned timescales (see the top panel 
of Fig.~\ref{fig:crosscor}). We also verified that no lag is observed when considering the softest energy range in the 
pn data sets (i.e. below 3~keV), and also when analysing the emissions separately in the different modes. On the other 
hand, the X-ray and UV emissions are not correlated over the whole duration of the observations (see bottom panel of 
Fig.~\ref{fig:crosscor}). Although a peak is visible at zero lag in the cross-correlation function, its significance 
is much lower than that of the peak measured between the soft and hard X-ray emissions, and comparable with that of 
other peaks (e.g. around $\sim-500$~s). These characteristics, as well as the overall asymmetric shape of the cross-correlation 
function, suggest that this peak is unlikely to reflect a true correlation. We note that some correlations and lags might be 
detected between X-rays and UV during the flares (as reported by Bogdanov et al. 2015 between the X-ray and optical bands). 
However, a multiband study of the system's flaring activity using our new data sets was precluded since \swift\ was not 
observing the source during the flaring episodes (see Fig.~\ref{fig:xmm_lcurve}).

\section{Spectral analysis}
\label{model}

\subsection{Data extraction for the new observations}

Since we detected no lags between the \xmm\ and the \nustar\ time series (see Sect.Fig.~~\ref{crosscor}), we created 
the good time intervals associated to each of the three modes by applying the above-mentioned intensity thresholds 
to the EPIC light curve, that is, that characterised by the highest counting statistics (see also Bogdanov et al. 2015; 
Campana et al. 2016), and applied them to all our \xmm\ and \nustar\ event files. The resulting exposure times and 
count rates for the X-ray modes and the different X-ray instruments are listed in Table~\ref{tab:states_log}.

For the spectral analysis of EPIC data we retained only single- and double-pixel events (\textsc{pattern} $\leq4$) 
for the pn, and single to quadruple-pixel events for the MOSs (\textsc{pattern} $\leq12$). Pixels and columns 
near the borders of the CCDs were excluded (\textsc{flag} = 0). Redistribution matrices and ancillary files were built 
with \textsc{rmfgen} and \textsc{arfgen}, respectively. For the \nustar\ data sets we extracted background-subtracted 
spectra and generated instrumental response files again separately for the two focal plane modules (FPMA and 
FPMB) with \textsc{nuproducts}. 

For each instrument, events were retained within the energy interval whereby the calibration of the spectral 
responses is best known, that is, 0.3--10~keV for the MOS, 0.7--10~keV for the pn (owing to known calibration uncertainties 
of this instrument at low energy in FT mode), and 3--79~keV for \nustar. Spectral channels were rebinned so as to contain 
a minimum of 200, 100, and 20 photons in each energy bin for the pn, MOS, and FPM data, respectively.

\begin{table}
\begin{center}
\caption{Log of the X-ray modes of J1023 from the simultaneous \xmm\ EPIC and \nustar\ data sets.}
\begin{tabular}{lccc}
\hline \hline
Mode                                    & Instrument    & Exposure      & Average count rate  \\
                                                &                       & (s)                     & (counts s$^{-1}$) \\
\hline
\multirow{5}{*}{\textit{Flare}}         & pn                    & 2613          & $13.94 \pm 0.07$        \\
                                                & MOS\,1                & 2533            & $3.92 \pm 0.04$       \\
                                                & MOS\,2                & 2400            & $3.83 \pm 0.04$       \\      
                                                & FPMA          & 1676          & $0.85 \pm 0.02$ \\
                                                & FPMB          & 1672          & $0.76 \pm 0.02$ \\
\hline                                                                                                                                                                          
\multirow{5}{*}{\textit{High}}          & pn                    & 52\,840               & $4.10 \pm 0.01$ \\ 
                                                & MOS\,1                & 55\,500         & $1.143 \pm 0.005$     \\
                                                & MOS\,2                & 51\,360         & $1.122 \pm 0.005$     \\      
                                                & FPMA          & 37\,870               & $0.208 \pm 0.003$       \\
                                                & FPMB          & 37\,660               & $0.200 \pm 0.002$       \\
\hline                          
\multirow{5}{*}{\textit{Low}}   & pn                    & 16\,210               & $0.596 \pm 0.009$ \\   
                                                & MOS\,1                & 17\,480         & $0.162 \pm 0.003$     \\
                                                & MOS\,2                & 15\,230         & $0.166 \pm 0.003$     \\      
                                                & FPMA          & 11\,570               & $0.023 \pm 0.002$ \\
                                                & FPMB          & 11\,560               & $0.027 \pm 0.002$ \\
\hline                                                                                  
\end{tabular}
\end{center}
{\bf Notes.} Count rates are in the same band as reported in Table~\ref{tab:log}.
\label{tab:states_log}
\end{table}

\subsection{Spectral fits for all high- and low-mode spectra} 
\label{analysis}

The spectral data sets of the new \xmm\ and \nustar\ observations, together with those extracted consistently from all archival \xmm\ and \axaf\ observations taken since 2008 
(see Campana et al. 2016) allow us to refine our model for the high-low mode switching in terms of transitions between the weak propeller and radio pulsar regimes\footnote{The 
bimodal shape of the X-ray flux distribution appears remarkably reproducible over a time span of years (2013--2017; see also Archibald et al. 2015; Bogdanov et al. 2015). 
Identical count rate thresholds were hence adopted to single out the high and low X-ray modes across all data sets.}. All spectra for the high and low modes were fitted together 
within \textsc{xspec} (Arnaud et al. 1996) with the same model described by Campana et al. (2016): for the high mode we adopted a power law component associated with 
the propelling magnetosphere (see Papitto \& Torres 2015); a radiatively inefficient advection-dominated accretion disc (described with the \textsc{diskpbb} model in \textsc{xspec}); 
a thermal component related to the X-ray pulsed emission arising from matter leaking through the propelling magnetosphere and accreting onto a restricted area of the NS surface 
(\textsc{nsatmos} in \textsc{xspec}; see Heinke et al. 2006). For the low mode we included a power law with $\Gamma \sim 2$ to model the emission at the shock front between 
the relativistic pulsar particle wind and matter inflowing from the companion (see Tavani \& Arons 1997), again a radiatively inefficient accretion disc, and two components related 
to the radio pulsar activity: a thermal one originating from the NS polar caps (expected to be only slightly colder than during the high mode owing to the short timescale of the mode 
switching), and a non-thermal, power law-like one, related to magnetospheric mechanisms. The absorption by the interstellar medium along the line of sight was described via the 
\textsc{TBabs} model (Wilms et al. 2000). 

In the spectral fits, we allowed the following parameters to vary: the disc temperature and inner radius, the polar cap temperature and radius, and the radio pulsar magnetospheric 
component. We required the disc temperature and inner radius to lie on the same $T(r)\propto r^{-p}$ curve across the two modes ($p$ was always kept fixed to 0.5), and the polar 
cap radius to be the same across the two modes and smaller than the NS radius. We tied up all parameters of the spectral components for the rotation-powered MSP state to those 
of the low mode. Finally, we tied up the hydrogen column density across all data sets. 

We obtained a statistically acceptable result ($\chi^2_{\nu}=1.05$ for 5246 d.o.f., after the addition of a systematic error of 2 percent; Smith 2015). We list the updated values for 
the parameters in Table~\ref{tab:spec}, and for plotting purpose we show only the simultaneous \xmm\ and \nustar\ spectra fitted with this model in Fig.~\ref{fig:xmm_states_spectra}, 
together with post-fit residuals.

\begin{table*}
\caption[X-ray spectral fits of J1023.]{Results of the spectral modelling for the X-ray high and low modes and the rotation-powered MSP state of 
J1023 (see the text for details).}
\label{tab:spec}
\begin{center}
\begin{tabular}{ccccc}
\hline \hline
Spectral parameter                                      & High mode                         & Low mode          & Rotation-powered MSP state\\
                                                        & (319.2~ks)                          & (104.6~ks)                       & (116.8~ks)             \\   
\hline \vspace{0.1cm}
power law $\Gamma_D$                    & $1.58\pm0.01$                  &$2.02\pm0.03$          & --                  \\ \vspace{0.1cm}
power law $N_D$ ($10^{-4}$)             & $17.6\pm0.4$                    & $2.87\pm0.06$                & --                  \\ \vspace{0.1cm}
disc $kT$ (eV)                                  & $123^{+18}_{-13}$      	& $<46$                             & --                  \\ \vspace{0.1cm}
disc norm. $N_d$                                & $161^{+155}_{-92}$          & $>8\times 10^3$  & --                   \\ \vspace{0.1cm}
NS atmos. $kT_\infty$  (eV)               & $99^{+9}_{-7}$                 & $50^{+5}_{-4}$              & tied to Low  mode   \\ \vspace{0.1cm}
NS atmos. $R_{em}$ (km)                 & $3.5^{+0.7}_{-0.6}$           & tied to High mode       & tied to High mode \\ \vspace{0.1cm}
NS power law $\Gamma_P$                 & --                                        & $1.02\pm0.05$              & tied to Low mode \\ \vspace{0.1cm}
NS power law $N_P$ ($10^{-5}$)  & --                                       & $3.5\pm0.3$                 & tied to Low mode \\ 
\hline
\end{tabular}
\end{center}
{\bf Notes.} The column density was tied up across all data sets, yielding $N_H = 5.5^{+0.5}_{-0.6} \times 10^{20}$~cm$^{-2}$.  
The fit provides $\chi^2_{\nu}=1.05$ for 5246 d.o.f., after the addition of a systematic error of 2 percent. \\
\noindent $^1$ Errors are quoted at the 90\% confidence level for one parameter of interest. \\
\noindent $^2$ The effective NS temperatures were corrected for the gravitational redshift: $kT_\infty = kT(1+z)^{-1}$ where  
$1+z = (1- 2GM_{NS}/R_{NS}c^2)^{-1/2}$ is the gravitational redshift factor, and $M_{NS}$ and $R_{NS}$ are the NS mass and radius.\\
\noindent $^3$ Radii were computed assuming a source distance of 1.37~kpc (Deller et al. 2012).\\
\end{table*}

\subsection{Spectral fits for the flaring mode spectra for the most recent observations} 
\label{analysis_flare}
 
\begin{figure}
\centering
\includegraphics[width=0.52\textwidth]{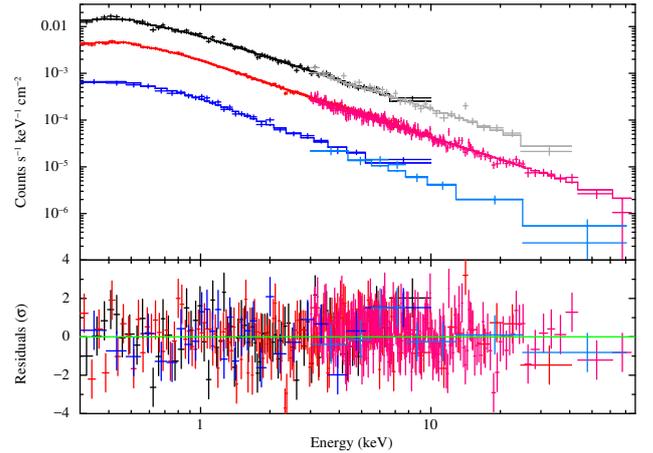}
\vspace{-20pt}
\caption{Broadband X-ray spectra of J1023 relative to the three X-ray modes from simultaneous \xmm\ and \nustar\ data. For plotting 
purposes, only the \xmm\ EPIC MOS1 (0.3--10~keV) and \nustar\ FPMA (3--79~keV) spectra are shown. Data have been re-binned to 
better visualise the trend in the spectral residuals. The best-fitting model is indicated by the solid lines. High- and low-mode spectra were 
fitted using the same model described by Campana et al. (2016; see Sect.~\ref{analysis}), whereas flare spectra were fitted separately 
with an absorbed power law model (see Sect.~\ref{analysis_flare}). Black and grey refer to the flaring mode, red and magenta to the 
high mode, and blue and cyan to the low mode. Post-fit residuals in units of standard deviations are also shown in the bottom panel. 
}
\label{fig:xmm_states_spectra}
\end{figure}

We have additionally fitted an absorbed power law model to the \xmm\ + \nustar\ spectra from the flaring mode, to detect possible variations in the spectral shape of this mode 
compared to that observed in previous observations. We fixed the column density to the value derived from the above-mentioned model, that  is, $N_H = 5.5 \times 10^{20}$~cm$^{-2}$ 
(see Table~\ref{tab:spec}) and obtained a photon index $\Gamma=1.67\pm0.01$ ($\chi^2_{\nu}=1.07$ for 255 d.o.f.). This is consistent with that observed up to 10~keV in the long \xmm\ 
observation in 2013 November (Bogdanov et al. 2015), implying no substantial variations in the spectral shape of the flaring mode over $\sim 2.5$ yrs. We show the \xmm\ 
and \nustar\ spectra with the best-fitting absorbed power law model in Fig.~\ref{fig:xmm_states_spectra}, together with post-fit residuals.

\section{High-resolution X-ray spectroscopy}
\label{sec:rgs_results}

\subsection{Model for the continuum}

Before searching for narrow features in the RGS first-order spectrum, we extracted the stacked background-subtracted spectra from the
MOS1 and MOS2 instruments from all observations carried out since the system transition to the LMXB state (see table~1 by Jaodand 
et al. 2016 for a complete journal of these observations), to increase the counting statistics and better characterise the model for the continuum 
emission. The pn spectra were not considered for this purpose, owing to the known uncertainties in the spectral calibration of the data sets 
acquired in the timing mode at low energy (typically below $\sim0.7$~keV). The MOS spectra were grouped to have at least 100 counts in 
each spectral bin.

We started by fitting the stacked MOS1, MOS2, and RGS spectra simultaneously with a power law model corrected for the absorption by 
the Galactic interstellar medium (\textsc{tbabs*powerlaw}), to adequately constrain the parameters for the continuum. The parameters were 
tied up across the three data sets, and a renormalization constant was included to account for intercalibration uncertainties. The choice of 
this simple phenomenological model to describe the continuum (compared to the more sophisticated one adopted for the different X-ray modes) 
was motivated by the fact that the power law component provides the largest contribution to the system's soft X-ray emission (see e.g. 
Fig.~3 by Campana et al. 2016). The counting statistics in the RGS spectrum is large enough to enable the use of the $\chi^2$ statistics to 
evaluate the goodness of the spectral fits (i.e. we verified that each spectral channel contained at least 20 photon counts). We obtained a 
photon index $\Gamma = 1.82\pm0.02$ and absorption column density $N_{\rm H} = 5.6_{-0.2}^{+0.3} \times 10^{20}$ cm$^{-2}$, which is 
fully compatible within the uncertainties with the value reported in Sect.~\ref{analysis}. 

\subsection{Characterisation of the narrow features}

\begin{figure*}
\centering
\includegraphics[width=.64\textwidth]{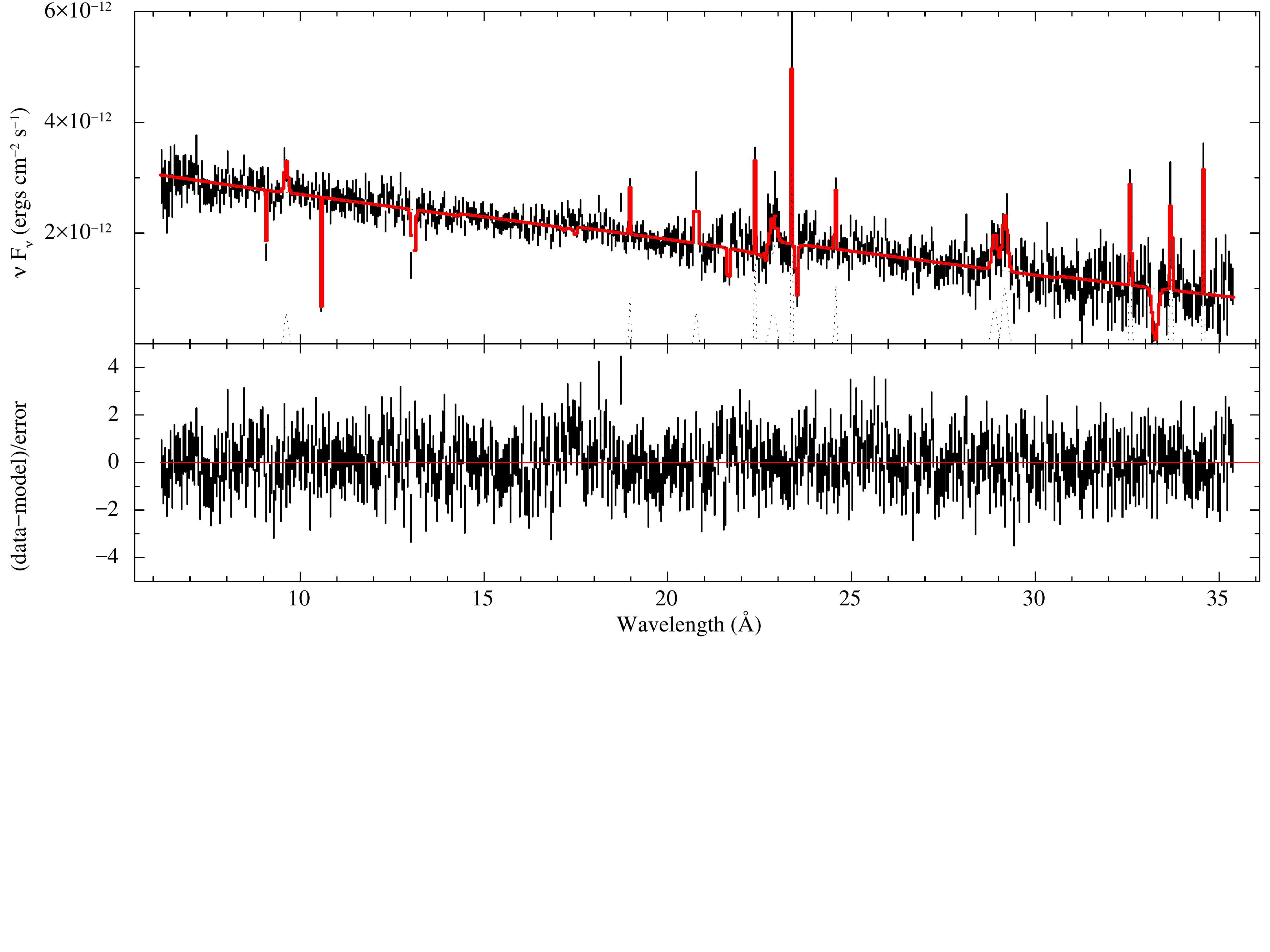}\\
\vspace{-2.9cm}
\includegraphics[width=.64\textwidth]{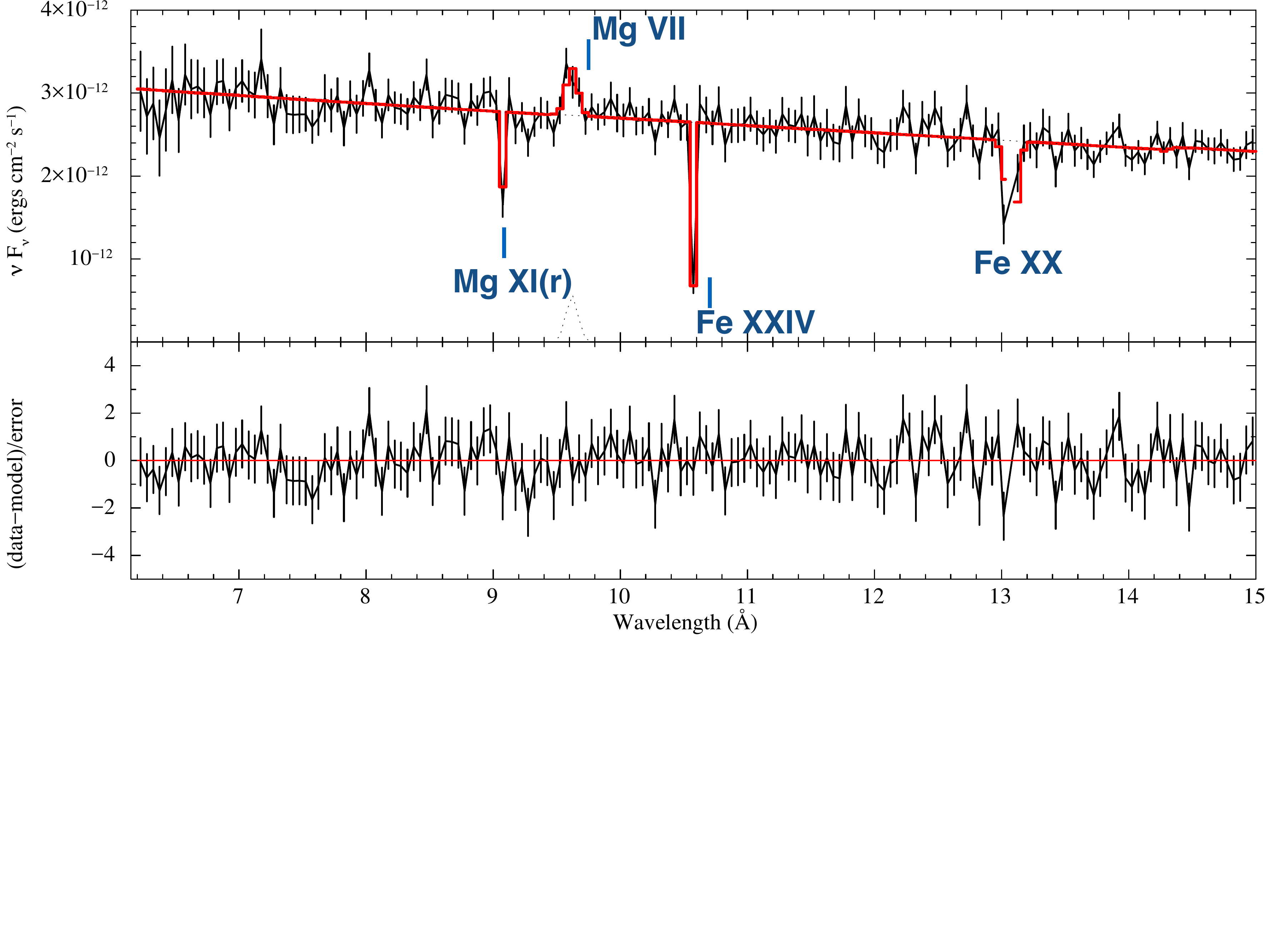}\\
\vspace{-2.9cm}
\includegraphics[width=.64\textwidth]{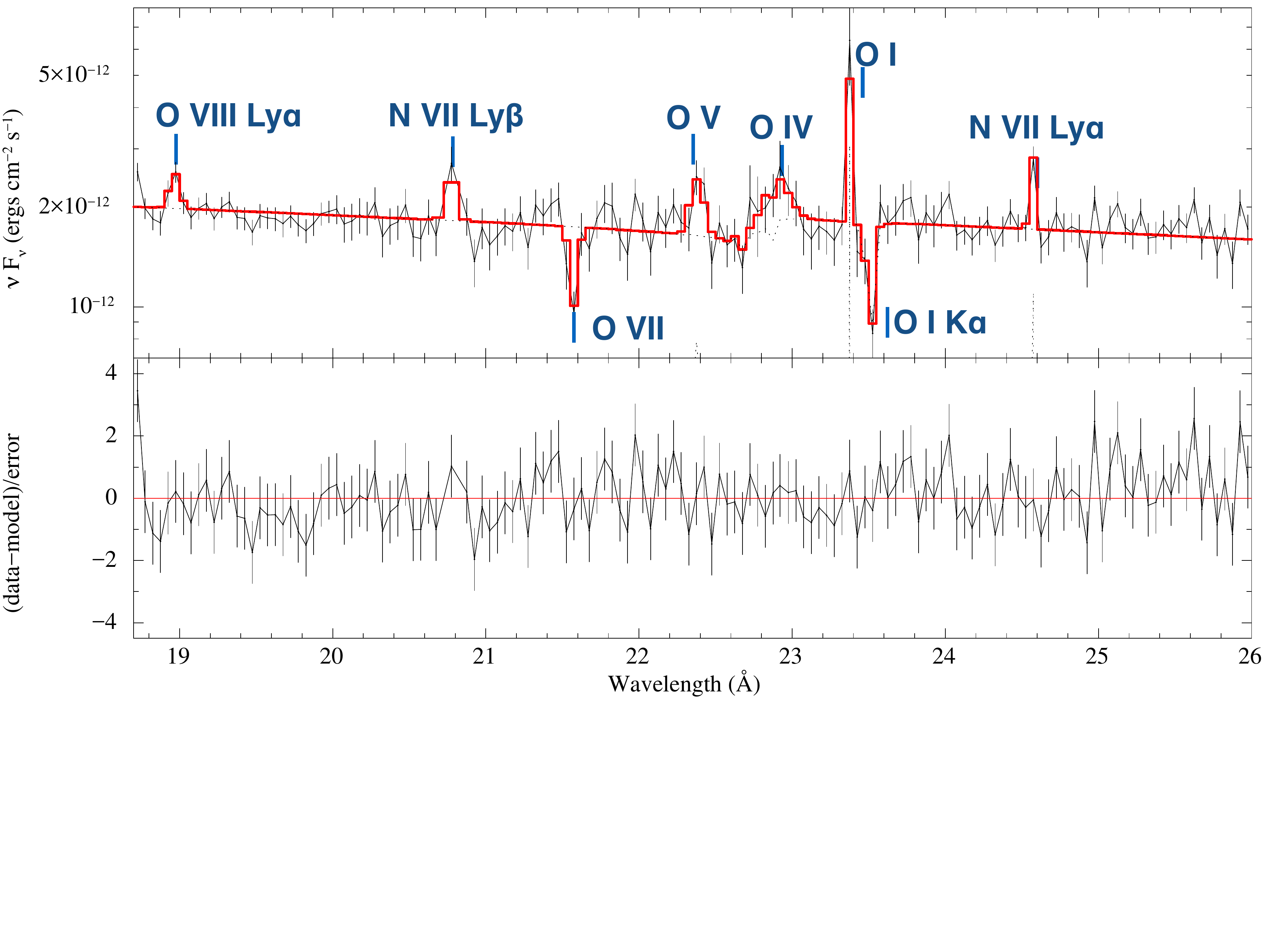}\\
\vspace{-2.3cm}
\includegraphics[width=.635\textwidth]{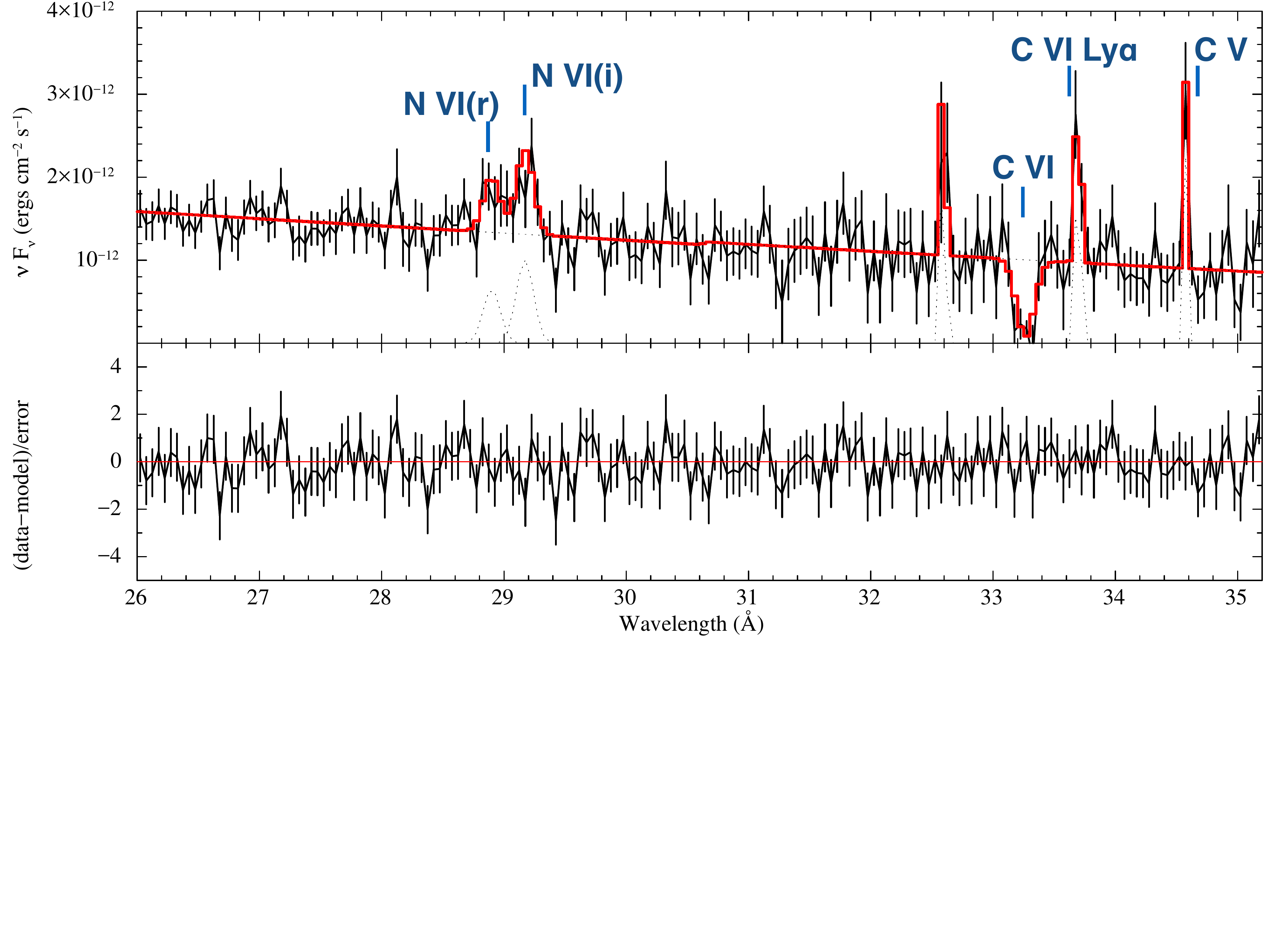}
\vspace{-2.9cm}
\caption{Unfolded RGS first-order spectrum of J1023 in the 6.2--35.4~\AA\ wavelength range (top panel) and in selected wavelength ranges (in order 
of increasing wavelength from top to bottom). The red solid line represents the best-fitting model (see the text for details). The wavelengths of the 
most relevant transitions are labelled in blue. Post-fit residuals are also shown.}
\label{fig:rgs_spectrum}
\end{figure*}

The poorer spectral resolution of the MOS instruments compared to the RGS precludes a detailed characterisation of narrow features in the 
EPIC data sets in the softest X-ray band. We hence focused on the RGS data, and fixed $N_{\rm H}$ and $\Gamma$ to the values derived 
from the simultaneous fit of the RGS and MOS data, that is, $N_{\rm H} = 5.6 \times 10^{20}$ cm$^{-2}$, $\Gamma = 1.82$. We visually inspected 
structured residuals with respect to the model for the continuum in the RGS spectrum, and fitted the most significant residuals by including 
Gaussian components to the model, either in emission or absorption (progressively and in order of decreasing wavelength). Several complex 
structures were observed in the wavelength range encompassing the oxygen transitions (in particular in the 21--24~\AA\ interval). We thus 
decided to substitute the \textsc{tbabs} model with the more physically motivated \textsc{tbnew}$_-$\textsc{feo} model, which represents an 
improved high-resolution version of the Tuebingen-Boulder model (Wilms et al. 2000) where the absorption column for the oxygen and the iron 
are allowed to vary in the fit. This choice allowed us to obtain a sound modelling of the oxygen K edge and other features related to oxygen 
transitions. In fact, this component self consistently accounts for the strong O I K$\alpha$ absorption line observed at 527.2~eV (see 
Fig.~\ref{fig:rgs_spectrum}). The abundances of the oxygen and iron were compatible with the solar values, and were hence fixed to those 
values. The widths of the Gaussian components were fixed to values smaller than the instrument energy resolution, that is, $\sigma=0$ or 1~eV.

The overall RGS spectrum can be satisfactorily described by including 16 Gaussian features (11 in emission and 5 in absorption). We obtained 
$\chi^2_\nu = 0.92$ for 538 d.o.f.. Table \ref{tab:rgs_results} reports the properties of the detected features as well as their identification with the 
transitions of the different elements. The expected energy of the transitions is also reported for comparison (see e.g. Engstr\"om \& Litz\'en 1995; 
Pradhan 2003; Gu et al. 2005; Smith \& Brickhouse 2014). The significance for each feature (either in emission or absorption) is reported in terms 
of the improvement in the value of the $\chi^2$ when including the corresponding Gaussian component to the model, and as the ratio between 
the line normalization and the associated uncertainty at the 1$\sigma$ confidence level. The uncertainty on the line normalization was evaluated 
by allowing both the line centroid and normalization to vary in the fit. We also verified that the inclusion of additional spectral components to describe 
the continuum (\textsc{nsatmos} and \textsc{diskpbb}) did not significantly affect the properties of the narrow features (in particular their centroids 
and equivalent widths) and the significance of detection.

\begin{table*}
\begin{center}
\caption{Results of the spectral fits of the stacked RGS first-order spectrum of J1023.} 
\label{tab:1023_rgs}
\begin{tabular}{lccccccc} 
\hline \hline
Element transition      & Wavelength    & Energy                                                & E$_{lab}$       & $|\Delta\lambda| / \lambda$           & EQW                           & $\Delta \chi^2$                         & Significance   \\
                                        & (\AA)         & (eV)                                                  & (eV)            &                                                       & (eV)                                    &                                               & ($\sigma$) \\ 
\hline          
 \vspace{0.1cm} 
C V 1s-3p                               & 34.56 & $358.7\pm0.5$  (e)                             & 354.5         & 0.01                                          & 1.3$^{+1.1}_{-0.9}$        & 4.62                                       & 2.0      \\  \vspace{0.1cm}   
C VI Ly$\alpha$                 & 33.69 & 368.0$^{+0.4}_{-1.6}$  (e)                             & 367.5         & 0.001                                         & 1.4$^{+0.8}_{-1.4}$        & 9.84                                       & 3.3  \\  \vspace{0.1cm} 
C VI Ly$\alpha$                 & 33.25 & $372.9\pm0.7$   (a)                            & 367.5                 & 0.01                                          & 1.7$^{+1.2}_{-1.2}$        & 7.53                                       & 3.1 \\    \vspace{0.1cm}    
N VI (i)                                & 29.19 & 424.7$^{+1.0}_{-0.7}$   (e)                    & 426.3         & 0.004                                         & 1.6$^{+0.8}_{-1.2}$        & 11.24                                      & 3.4  \\  \vspace{0.1cm} 
N VI (r)                                & 28.87 & 429.4$^{+1.5}_{-1.7}$   (e)                    & 430.7         & 0.003                                         & 1.2$^{+1.1}_{-0.8}$        & 6.01                                       & 2.7      \\  \vspace{0.1cm} 
N VII Ly$\alpha$                & 24.56 & 504.8$^{+0.6}_{-2.0}$   (e)                    & 500.3         & 0.009                                                 & 0.7$^{+0.9}_{-0.4}$      & 5.28                                         & 2.9  \\  \vspace{0.1cm} 
O II 1s-2p                      & 23.38 & 530.4$^{+0.6}_{-0.3}$   (e)                    & 532.9                 & 0.005                                                 & 1.9$^{+0.8}_{-1.2}$        & 2.69                                       & 2.1  \\  \vspace{0.1cm} 
O IV 1s-2p                      & 22.89 & 541.7$^{+2.3}_{-2.1}$   (e)                    & 543.06                & 0.003                                         & 1.6$^{+1.1}_{-0.8}$        & 4.84                                       & 2.3  \\  \vspace{0.1cm} 
O V 1s-2p                               & 22.38 & 554.0$^{+1.7}_{-1.3}$   (e)                    & 552.29                & 0.003                                         & 1.2$^{+0.7}_{-0.7}$        & 7.60                                       & 2.7 \\  \vspace{0.1cm} 
O VII 1s-2p                     & 21.57 & 574.7$^{+0.9}_{-0.5}$    (a)                   & 573.95                & 0.001                                         & 0.8$^{+0.3}_{-0.4}$        & 1.9                                        & 1.9 \\  \vspace{0.1cm} 
Fe XX                                   & 13.04 & 950.5$^{+1.2}_{-1.8}$    (a)                   & 949.3         & 0.001                                         & 4.4$^{+2.0}_{-2.2}$        & 16.95                                      & 3.9  \\  \vspace{0.1cm} 
Fe XXIV                         & 10.56 & 1173.8$^{+2.1}_{-2.0}$   (a)                   & 1167.6                & 0.005                                         & 4.1$^{+1.2}_{-1.7}$        & 21.52                                      & 4.6 \\   \vspace{0.1cm}   
Mg VII 1s-2p                    & 9.60          & $1291.0\pm4.1$   (e)                           & 1291.6                & 0.0005                                                & 3.2$^{+1.2}_{-1.7}$        & 10.44                                      & 2.8 \\  \vspace{0.1cm} 
Mg XI (r)                               & 9.07          & 1366.7$^{+1.7}_{-3.4}$  (a)                    & 1357          & 0.007                                         & 2.9$^{+1.8}_{-1.9}$        & 5.05                                       & 2.3 \\ 
\hline
\label{tab:rgs_results}
\end{tabular}
\end{center}
{\bf Notes.} Letters in parentheses indicate whether the narrow features are detected in emission (e) or absorption (a). EQW denotes the equivalent width of the Gaussian feature. Errors are reported at 
the 90 percent confidence level for a single parameter of interest. $\Delta \chi^2$ indicates the improvement in the overall fit when adding the Gaussian component to the model.
\end{table*}

The spectrum with the best-fitting model superimposed is shown in Figure \ref{fig:rgs_spectrum} (both in the overall RGS wavelength coverage, 
as well as in selected wavelength ranges where the most prominent lines are detected). The H-like lines from C VI, N VII, and O VIII, as well as 
the C V, O II, O IV, O V, Fe XX, Fe XXIV and Mg VII transitions are all detected close to, or compatible with, their expected rest frame energies or 
wavelengths. The line centroid energies measured for the N VI triplet of 424.7$^{+1.0}_{-0.7}$~eV and 429.4$^{+1.5}_{-1.7}$ are consistent with 
the majority of the emission originating from the intercombination (near 426.3~eV) and resonance (430.7~eV) transitions, respectively, but are 
inconsistent with the expected energy of the forbidden (at 419.8~eV) emission line. On the other hand, the line centroid energy measured at 
1366.7$^{+1.7}_{-3.4}$~eV is close to the energy of the Mg XI resonance transition at an expected energy of 1357~eV.

We then tested whether it was possible to account for most of the emission lines using a more physically motivated collisionally ionized plasma model. 
We replaced the Gaussian lines with either a \textsc{mekal} or \textsc{apec} component, as well as with the absorption grids 18 and 21 from the 
\textsc{xstar} library (Kallman et al. 1996)\footnote{See https://heasarc.gsfc.nasa.gov/lheasoft/xstar/xstar.html.}. These cover similar ranges in both the 
ionization parameter (log$\epsilon$ from -4 to +4 ergs cm s$^{-1}$) and column densities ($N_H$ from 10$^{19}$ to 10$^{23}$~cm$^{-2}$), but assume 
significantly different turbulent velocities, 100 and 1000 km~s$^{-1}$ for the grid 18 and 21, respectively. Statistically acceptable fits were obtained in both 
cases by allowing the abundances of the different elements to vary ($\chi^2_\nu = 1.05-1.10$ for 577 d.o.f.). However, none of these models were 
able to reproduce properly the narrow features: visual inspection of the spectrum reveals structured residuals around most of the features. More complicated 
models are beyond the scope of this work.

We also performed an orbital phase-resolved spectral analysis of the system. We used the ephemerides calculated by Jaodand et al. (2016) to extract 
for each dataset the event files and spectra in three orbital phase bins of equal width (0--0.33, 0.33--0.66 and 0.66--1). We then merged all data sets 
corresponding to the same phase bins. In the spectral analysis, all line parameters, the column density, and the power law photon index were kept fixed 
to their phase-averaged values, whereas the power law normalization was left free to vary to account for possible variations of the X-ray flux as a function 
of the orbital phase. We obtained acceptable fits in all cases, with $\chi^2_\nu$ ranging from 0.9 to 1.0. Visual inspection of the spectrum in restricted 
wavelength ranges revealed that the renormalized model provided an accurate modelling of the line shapes. We conclude that there are no statistically 
significant variations of the line properties along the orbital phase.

\section{Discussion and conclusions}
\label{discussion}

We report on the results of a simultaneous multi-wavelength observational campaign of the transitional millisecond pulsar J1023, encompassing the X-ray (\xmm\ EPIC 
and \nustar), UV (\swift), and optical (\xmm\ OM) bands. The deep ($\sim120$~ks) \xmm\ observation revealed the expected trimodal behaviour of J1023 (i.e. 
high, low, and flaring mode) in its current state, showing intensity levels and timescales of switching remarkably similar to those observed in all the \xmm\ data 
sets of the system acquired in the past 4 years. The \nustar\ data show a completely consistent variability at higher energy and on the same timescales (see 
Fig.~\ref{fig:xmm_lcurve}). 

The simultaneous \xmm\ and \nustar\ observations of J1023 presented in this work allowed us to probe possible spectral hardening of the X-ray emission on an extended 
energy range. We do not observe significant spectral changes with energy in the different modes. J1023 shows some similarities with the other well-known transitional 
millisecond pulsar XSS\,J12270$-$4859 in the LMXB state. The detailed broad-band studies conducted by de Martino et al. (2010, 2013) indeed revealed that 
XSS\,J12270$-$4859 also does not show evidence for conspicuous spectral variability among the three modes, except for some hardening especially during the 
low-mode episodes following the flaring events.

Our multi-wavelength campaign allowed us to search for correlated variability and lags of the system emission at different wavelengths, including for the first time, the hard X-ray 
energy range. The cross-correlation function between the soft X-ray and hard X-ray emissions is characterized by a remarkably symmetric shape, and we do not observe any lag 
on the sampled timescales. Although the uneven sampling of \nustar\ and the lower counting statistics available especially at high energy precluded detailed characterisation 
of the shape of the cross-correlation function during the different modes, inspection of the light curves binned at 10~s revealed a completely consistent behaviour in the soft and 
hard X-ray bands at the different luminosity levels. These results suggest that the bulk of the X-ray emission arises from the same physical mechanism. According to our model 
(see Campana et al. 2016), this should be associated to synchrotron radiation at the boundary between the propelling magnetosphere and the disk-mass in-flow in the high mode 
(see Papitto \& Torres 2015); synchrotron radiation in the local magnetic field at the shock front between the pulsar particle wind and the intra-binary material in the low mode 
(with negligible thermal contributions from the accretion disc and the pulsed emission from the NS surface during both modes; see Fig.~3 of Campana et al. 2016). Interestingly, 
de Martino et al. (2010, 2013) computed the cross-correlation function for the TMP XSS\,J12270$-$4859 in different soft X-ray (<10~keV) bands during the LMXB state, and found 
a symmetric shape of the correlation function with no detectable time lag during both the flaring and low modes. These results are similar to what we report here for J1023 over 
a broader energy range (up to 79~keV) and also for the case of the high mode. 

On the other hand, we do not observe any significant correlation between the X-ray and UV emissions during the high and low modes. This is consistent with the results reported 
by Bogdanov et al. (2015), who performed cross-correlations between the X-ray and optical bands using two long \xmm\ observations in 2013 November and 2014 June and found 
no correlation during the two modes. This result suggests a different origin for the bulk of the emission in the UV/optical bands, likely from the asymmetrically heated face
of the companion star (Papitto et al., in preparation; note that the optical pulsed emission discovered by Ambrosino et al. 2017 plays only a minor role in the optical budget, as it shows 
a fractional amplitude always below 1 percent). Unfortunately, the X-ray flares observed during our campaign were not covered by \swift\ UVOT, precluding a comparison between 
the X-ray and UV emissions during the flaring mode. We note that Bogdanov et al. (2015) found a significant correlation for the flaring mode, with some flaring episodes preceding 
those in the X-rays and vice versa, and having time delays of a few hundred seconds. 

The new observations provide an updated characterisation of the high- and low-mode X-ray spectra, and confirm and constrain the main properties of our model (see Campana et al. 
2016). In particular, we found a smaller variation of the disc black body component normalization across the high and low modes with respect to our previous data modelling (see 
Table~\ref{tab:spec}). This parameter can be related to the magnetospheric radius, at which the pressure of the NS magnetosphere equals the ram pressure of the inflowing matter. 
We found that in the high mode, the magnetospheric radius attains a value $r_m = 26^{+11}_{-9}$~km, whereas in the low mode only a lower limit $r_m>185$~km can be inferred. 
It is typically assumed that the inner edge of the accretion disc is truncated at the magnetospheric radius. Therefore, in the high mode, the inner disc radius is fully consistent within 
the uncertainties with lying between the corotation radius $r_{cor} = (G\,M\,P^2/4\,\pi^2)^{1/3} \sim 24$~km, at which matter in Keplerian orbit corotates with the NS, and the light 
cylinder radius $r_{lc} = c\,P/2\,\pi \sim 81$~km, at which a particle corotating with the magnetosphere would move at the speed of light (here $P$ is the NS spin period and $c$ the 
speed of light). In the low mode, the inner disc radius is instead pushed beyond the light cylinder radius. In the framework of the basic theory for the accretion regimes in NS LMXBs
(see e.g. Campana et al. 1998), these results further corroborate our scenario where the repeated transitions between the two modes are due to fast switches between a weak propeller 
regime in the high mode ($r_{cor}\lesssim r_m <r_{lc}$) and the radio pulsar mechanism in the low mode ($r_m > r_{lc}$). The observed radio brightening associated with the X-ray 
low mode (Bogdanov et al. 2017) can be easily reconciled with our model, because an enhancement of the radio emission is naturally expected when the inflowing matter is rapidly 
ejected by the radiation pressure of the pulsar.

The transitions between the different accretion regimes onto MSPs were recently explored by Parfrey \& Tchekhovskoy (2017), who performed the first time-dependent general-relativistic
magnetohydrodynamics simulations of the accretion of magnetized plasma onto a rotating, magnetized NS. Interestingly, their simulations produce a peculiar regime where the accretion flow 
spends most of the time inside the light cylinder (i.e. the propeller regime), but is regularly expelled out through it (i.e. the radio pulsar ejection regime via the pulsar wind). The simulations 
show that these transitions can take place over a few seconds (K. Parfrey, priv. comm.). This theoretical result is again fully compatible with our scenario for the mode switching 
of J1023, with the fact that the system is observed most of the time in the high mode (propeller) and only a minority of the time in the low mode (particle ejection in the form of pulsar wind), and 
with the short timescale observed for the mode switching. 

The fractional amplitude of the optical pulsations recently discovered from the system showed variations in time up to a factor of $\sim4$ on timescales of $\sim1.2$~ks, and upper limits were 
reported during a few time spans (Ambrosino et al. 2017; see in particular their Fig.~2). The lack of simultaneous X-ray observations prevented the authors from clarifying if the 
appearance/disappearance of the optical pulses is strictly related to the mode switching. According to our scenario, optical pulsations should be observed only in the low mode if these are 
due to synchrotron emission by relativistic electrons and positrons in the magnetosphere of a rotation-powered pulsar, as proposed by Ambrosino et al. (2017). On the other hand, detection 
of optical pulsed emission only in the high mode (i.e. simultaneously with the X-ray pulsations) would require that a different emission mechanism is at work, possibly related to residual 
channelled accretion onto the NS polar caps in the weak propeller regime. The maximum pulsed luminosity achievable in the optical band via cyclotron emission by thermal electrons located 
within the accretion columns, $L_{cycl}\sim3 \times 10^{29}$ \lum, is smaller than the maximum observed pulsed luminosity, $L_{opt,puls}\sim10^{31}$ \lum, only by a factor of $\sim30$ 
(Ambrosino et al. 2017). A beamed cyclotron emission pattern with beaming factor $b = L_{cycl} / L_{opt,puls} \sim 1/30$ cannot be ruled out a priori, and might be in principle a viable possibility 
(assuming that the maximum optical pulsed luminosity is attained during the high mode; otherwise $b$ might be even larger). If instead optical pulsations are detected both in the low and high 
modes and they arise indeed from the two above-mentioned different processes, we shall expect to observe some differences in the pulse profile shape and amplitude across the two modes 
in the optical. Alternatively, an intriguing possibility is that a radio pulsar dipole spin-down mechanism might be constantly at work (as already suggested by Coti Zelati et al. 2014; Stappers et al. 
2014; Takata et al. 2014), regardless of the source state, converting a tiny fraction ($\sim2\times10^{-5}$) of the spin-down power into optical pulsed emission. 
In all the above-mentioned scenarios, the simplest explanation for the non-detection of the rotation-powered radio pulsations would be that these are overwhelmed by the accreting matter during 
the propeller regime in the high mode, and enshrouded by a large amount of intervening ionized material around the NS during the particle ejection regime in the low mode. Simultaneous X-ray 
and optical observations with high time resolution and large enough photon counting statistics are necessary to single out the different modes of J1023 in the optical band, characterize accurately 
the optical pulsations as a function of the luminosity modes, and test our predictions.  

We also performed a detailed high-resolution X-ray spectroscopy of the system using all RGS data acquired on J1023 since it became much brighter in the X-rays in 2013. This allowed us 
to identify several narrow features in the soft X-ray spectrum. A commonly used diagnostic for the emitting regions can be derived from the ratio between the different components of the 
Helium-like triplets (Porquet \& Dubau 2000). In particular, a first-order estimate of the density of the emitting gas can be inferred from the parameter $R = z/(x+y)$, which quantifies the ratio 
between the intensity of the forbidden line ($z$) and that of the sum of the intercombination emission ($x+y$). For the case of J1023, the best constraint on $R$ is provided by the N VI triplet, 
which shows a prominent intercombination emission but lacks the forbidden line. Assuming that the emission lines originate in a pure photo-ionized plasma, this implies a high-density plasma, 
with electron density $n_e \gtrsim 10^{11}$~cm$^{-3}$ (see e.g. Fig.~8 of Porquet \& Dubau 2000). This lower limit is consistent with the emission lines being produced within the accretion disc.

A detailed study of possible correlations and lags between the UV, optical, and near infrared emissions on timescales of seconds was impossible, owing to a non-optimal setup of the \xmm\ OM 
and bad weather conditions that impeded ground-based observations. However, this analysis will be the object of future work thanks to approved and ongoing multi-wavelength campaigns.

\begin{acknowledgements}
We thank Kyle Parfrey and Amruta Jaodand for helpful discussions, and Alessandro Papitto and Domitilla de 
Martino for comments on the manuscript. We thank the referee for useful comments and suggestions. We acknowledge 
the International Space Science Institute (ISSI) that funded and hosted the international team `The disk-magnetosphere 
interaction around transitional millisecond pulsars', and we thank all the members of the team for fruitful discussions. 
FCZ acknowledges Universit\`a dell'Insubria and the Anton Pannekoek Institute for Astronomy of the University of 
Amsterdam for their kind hospitality during the time that some of this work was carried out. The scientific results reported 
in this study are based on observations obtained with \xmm\ and \nustar. \xmm\ is an ESA science mission with 
instruments and contributions directly funded by ESA Member States and the National Aeronautics and Space Administration 
(NASA). The \nustar\ mission is a project led by the California Insitute of Technology, managed by the Jet Propulsion 
Laboratory, and funded by NASA. 
The \nustar\ Data Analysis Software (NuSTARDAS) is jointly developed by the ASI Science Data Center (ASDC, Italy) and the California Institute 
of Technology (Caltech, USA). 
FCZ, NR and DFT are supported by grants AYA2015-71042-P 
and SGR2014-1073. SC acknowledges support from the ASI/NuSTAR NARO16 funds. NR acknowledges funding in the framework 
of the Netherlands Organization for Scientific Research (NWO) Vidi award number 639.042.321.  
\end{acknowledgements}

\end{document}